\documentclass[aps, prb, onecolumn,groupedaddress]{revtex4} %ref style ^1
\usepackage{graphicx}
\usepackage{epstopdf}
\usepackage{amsmath}
\usepackage{amssymb}
\usepackage{ifthen}
\usepackage{booktabs}
\usepackage{gensymb}
\usepackage{color}

\newcommand{\bb}{\begin{equation}}
\newcommand{\ee}{\end{equation}}
\newcommand{\ba}{\begin{eqnarray*}}
\newcommand{\ea}{\end{eqnarray*}}
\newcommand{\rhor}{\rho({\bf r})}
\newcommand{\dd}{{\rm d}}
\newcommand{\rr}{{\mathbf r}}
\newcommand{\dr}{{\rm d}{\bf r}}

\usepackage{graphicx}% Include figure files
\usepackage{dcolumn}% Align table columns on decimal point
\usepackage{bm}% bold math
\usepackage{longtable}

\begin{document}

%\preprint{}

\title{Filling transitions in acute and open wedges}

\author{Alexandr \surname{Malijevsk\'y}}
\affiliation{
%{E. H{\'a}la Laboratory of Thermodynamics, Institute of Chemical Process Fundamentals, Academy of Sciences, 16502 Prague 6, Czech Republic}\\
{Department of Physical Chemistry, Institute of
Chemical Technology, Prague, 166 28 Praha 6, Czech Republic;}\\
 {Institute of Chemical Process Fundamentals, Academy of Sciences, 16502 Prague 6, Czech Republic}}
\author{Andrew O. \surname{Parry}}
\affiliation{Department of Mathematics, Imperial College London, London SW7 2BZ, UK}

\begin{abstract}

We present numerical studies of first-order and continuous filling transitions, in wedges of arbitrary opening angle $\psi$, using a microscopic fundamental measure
density functional model with short-ranged fluid-fluid forces and long-ranged wall-fluid forces. In this system the wetting transition characteristic of the planar
wall-fluid interface is always first-order regardless of the strength of the wall-fluid potential $\varepsilon_w$. In the wedge geometry however the order of the
filling transition depends not only on $\varepsilon_w$ but also the opening angle $\psi$. In particular we show that even if the wetting transition is strongly
first-order the filling transition is continuous for sufficient acute wedges. We show further that the change in the order of the transition occurs via a tricritical
point as opposed to a critical-end point. These results extend previous effective Hamiltonian predictions which were limited only to shallow wedges.

\end{abstract}

%Use showkeys class option if keyword

%\maketitle must follow title, authors, abstract, \pacs, and %\keywords
\maketitle

\section{Introduction}

Fluid interfacial phenomena have been a topic of intense research within statistical physics for over three decades -- see for example the review articles
\cite{evans79, rw, croxton, dietrich, henderson92,evans90, binder, bonn, saam}. It is now well established that confining a fluid or, equivalently, placing it in a
strong external potential may induce novel phase transitions, scaling behaviour and criticality distinct from those of the bulk fluid. Well studied examples of this
include wetting transitions at planar walls \cite{sullivan_gama, schick, dietrich, bonn} and capillary condensation in parallel plate geometries \cite{evans86,
evans87} and capillary grooves \cite{evans_cc, roth, our_groove, het_groove}. For wetting phenomena  it was quickly established that the order of the transition
depends very sensitively on the range and strength of the competing wall-fluid and fluid-fluid forces. In particular, in three dimensions, continuous (critical)
wetting transitions require, in general, a fine tuning of the interaction strengths, meaning that in experiments and in model calculations alike, the transition is
most often first-order in character. On non-planar substrates, however, geometrical effects are also important, and may strongly influence the order of any phase
transition. This is well illustrated by the filling transition occurring for a fluid confined in a linear wedge formed by two planar walls meeting at an opening angle
$\psi$. In many respects the filling transition is a missing link between wetting and capillary condensation, connecting these apparently two distinct phenomena, as
well as showing several novel features. Macroscopic arguments dictate, that a wedge in contact with a bulk vapour at two phase coexistence, is completely filled by
liquid if the contact angle $\theta<\theta_f$ satisfies \cite{concus, hauge, rejmer}
 \bb
 \theta_f(T)=\frac{\pi-\psi}{2}\,.\label{fill}
 \ee
where $(\pi-\psi)/2$ is often referred to as the wedge tilt angle $\alpha$. The filling transition  corresponds to the divergence in the adsorption as
$\theta(T)-\alpha\to 0^+$ and can be induced either at fixed $T$ by increasing the tilt angle (i.e. making the wedge more acute), or by increasing $T\to T_f$ causing
the contact angle to decrease until the condition $\theta(T_f)=\alpha$ is fulfilled. Wedge filling therefore precedes any wetting transition in the sense that
$T_f<T_w$ where $T_w$ is the wetting temperature at which $\theta(T_w)=0$. Indeed wedge filling does not actually require there to be any underlying substrate wetting
transition since it needs only the familiar phenomena of partial wetting. This makes filling transitions easier to observe than wetting since one need only tune the
substrate geometry rather than the details of the intermolecular interactions required to make the contact angle vanish. On the other hand, assessing the order of the
filling transition is more difficult than for wetting since there is no analogue of the macroscopic contact angle whose measurement would indicate the order of the
phase transition. However, the order of filling transitions can be distinct from that of wetting and is key to understanding the more subtle aspects of the phase
transition. These include strongly enhanced interfacial fluctuations and, in two dimensions, hidden connections with critical wetting referred to as wedge covariance
\cite{wood1, wood2, wood3}. All of these predictions arose initially from studies based on very simple effective Hamiltonian models which generalized the standard
Capillary-Wave analysis of wetting transitions to the wedge geometry. More recently however these predictions have been tested using microscopic models both at
mean-field level and beyond. For example in two dimensions, in addition to the known solution for wedge filling in the square lattice Ising model at a right-angle
corner \cite{abraham02, abraham03}, the transition has been studied within a field theoretical continuum model of fluid phase coexistence, which admits an exact
solution for arbitrary opening angles \cite{delfino}. This has shown that the wedge covariance has a deeper relation to the Lorentz invariance of quantum field models
in $1+1$ dimensions. In addition in three dimensions it has been possible to test predictions for the order of filling transitions in simulations \cite{binder03,
binder05} and using square-gradient theory \cite{bernardino} and modern density functional models based on fundamental measure theory \cite{our_prl, our_wedge}. This
has illustrated, that for right angle wedges the filling transition may indeed be continuous even though the underlying wetting transition is first-order.
Interestingly the mechanism for this change in order appears to be even more general than originally thought based on simple effective interfacial models.

\begin{figure}[h]
\includegraphics[width=4cm]{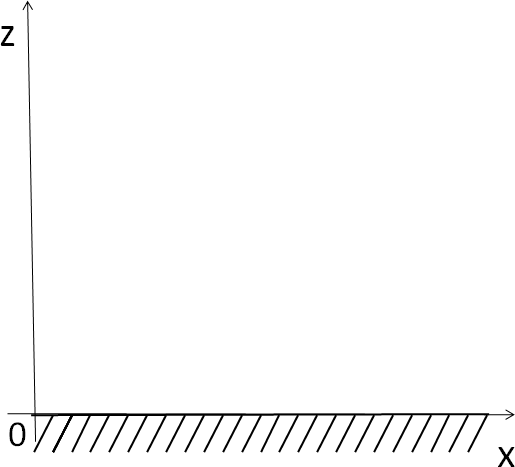}\hspace*{2cm}\includegraphics[width=4cm]{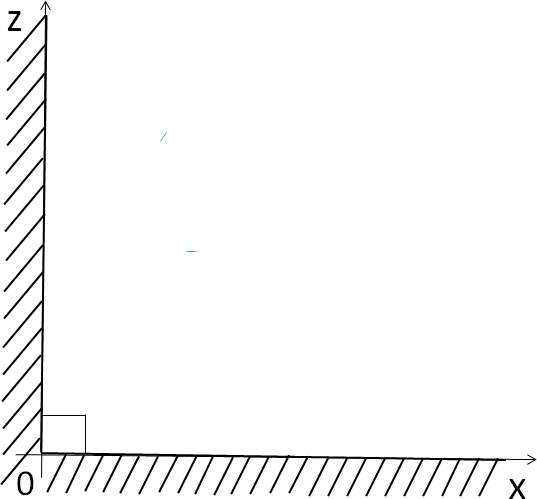}

 \vspace*{1cm}

\includegraphics[width=6cm]{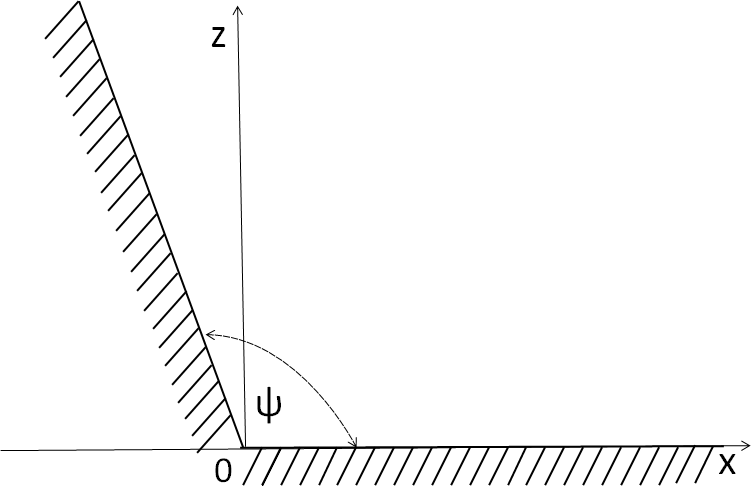}\hspace*{2cm}\includegraphics[width=5cm]{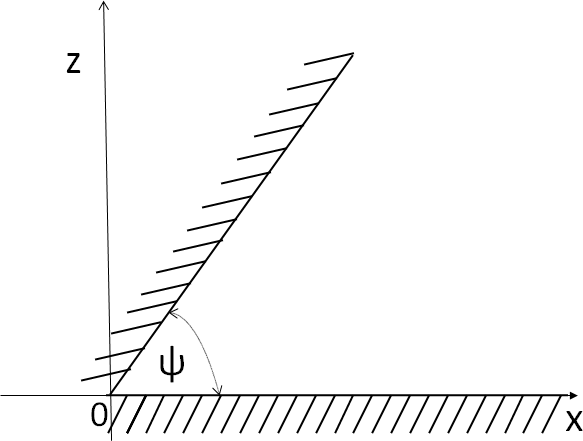}
\caption{Schematic illustrations of geometrical cross-sections for a) planar wall, b) rectangular wedge, c) open wedge, d) acute wedge. Translation invariance is
assumed along the $y$-axis. } \label{fig1}
\end{figure}

The purpose of the present work is to extend our recent density functional studies of wedge filling at right angle corners to more general opening angles. In
particular we wish to show that filling transitions that were observed to be first-order (continuous) for a right angle wedge, may become continuous (first-order) by
making the opening angle smaller (larger). Our article is arranged as follows: In the next section we present our microscopic model and the derivation of a
long-ranged external potential arising from dispersion-like forces, for a three dimensional wedge with arbitrary opening angle. We then consider a right-angle corner
with two different strengths of the wall-fluid potential, which give first-order and continuous filling respectively. These are then studied for different opening
angles showing the change in order as the wedge is made more acute and open, respectively. We conclude our paper with a discussion of the mechanism regarding the
change in the order of the transition.

\section{Density functional theory}

Within classical density functional theory the equilibrium density profile $\rhor$ is obtained from minimization of the grand potential functional
 \bb
 \Omega[\rho]={\cal{F}}[\rho]+\int\dr(V(\rr)-\mu)\rhor\,.\label{om}
 \ee
where $V(\rr)$ is the external field and $\mu$ is the chemical potential. All the information about the fluid model is contained in the intrinsic free energy
functional ${\cal{F}}[\rhor]$ which is often split into an ideal gas and excess contribution. Thus,
 \bb
 {\cal{F}}[\rho]={\cal{F}}_{\rm id}[\rho]+{\cal{F}}_{\rm ex}[\rho]\,,
 \ee
where ${\cal{F}}_{\rm id}[\rho]=k_BT\int\dr\rhor\left[\ln(\Lambda^3\rhor)-1\right]$ and $\Lambda$ is the thermal de Broglie wavelength that can be set to unity
without loss of generality.

In the spirit of van der Waals theory, the excess term is treated in a perturbative manner, and is separated into a) a contribution modelling the repulsive
hard-sphere (hs) core and b) a contribution from the attractive part $u(r)$ of the fluid-fluid intermolecular potential which is treated in simple mean-field fashion.
Hence we write
 \bb
 {\cal{F}}_{\rm ex}[\rho]={\cal{F}}_{\rm hs}[\rho]+\frac{1}{2}\int\dr\rhor\int\dr'\rho(\rr')u(|\rr-\rr'|)\,.\label{fex}
 \ee
where, in our analysis, $u(r)$ is taken to be a truncated Lennard-Jones-like potential
 \bb
 u_{\rm a}(r)=\left\{\begin{array}{cc}
 0\,;& r<\sigma\,,\\
-4\varepsilon\left(\frac{\sigma}{r}\right)^6\,;& \sigma<r<r_c\,,\\
0\,;&r>r_c\,.
\end{array}\right.
 \ee
which is cut-off at $r_c=2.5\,\sigma$. The hard-sphere term ${\cal{F}}_{\rm hs}[\rho]$  describes the repulsion between the fluid particles of diameter $\sigma$ which
is approximated using Rosenfeld's fundamental measure theory  as \cite{ros}
 \bb
{\cal{F}}_{\rm hs}[\rho]=\int\dr\Phi(\{n_\alpha\})\,.\label{fhs}
 \ee
where the $\{n_\alpha\}$ are six weighted densities.

In general, the external potential $V(\rr)$ for arbitrary wall shapes can be constructed by integrating a two-body wall-fluid potential $\phi_w(r)$ over the volume
$\cal{V}$ of the wall which is assumed to be of uniform density $\rho_w$:
 \bb
 V(\rr)=\rho_w\int_{\cal{V}}\dr' \phi_w(|\rr-\rr'|)\label{pot_gen}\,,
 \ee
 where in our study $\phi_w(r)$ is taken to be
 \bb
 \phi_w(r)=-4\varepsilon_w\left(\frac{\sigma}{r}\right)^{6}; \hspace{1cm}r>\sigma\,. \label{phi}
 \ee
In addition we impose a hard wall repulsion $V(\rr)=\infty$ whenever the distance from surface of the wall is less than $\sigma$. We work always in three dimensions
but assume translational invariance along the $y$ axis, so that the potential is only a function of the Cartesian coordinates $x$ and $z$ (see illustrations in
Fig.~1).

For the simplest case of a planar wall occupying a half space $z<0$ (see Fig.~1a), the potential reduces to a pure one-dimensional power-law
 \bb
 V_\pi(x,z)=\frac{2\alpha_w}{z^3}\,;\;\;\;z>\sigma \label{pot_pla}
 \ee
where $\alpha_w=-\frac{1}{3}\pi\varepsilon_w\rho_w\sigma^6$ measures the strength of the interaction. Similarly, for a right angle corner (Fig.~1b) the potential
maybe written \cite{our_prl}
 \bb
 V_{\pi/2}(x,z)=\alpha_w\left[\frac{1}{z^3}+\frac{2z^4+x^2z^2+2x^4}{2x^3z^3\sqrt{x^2+z^2}} +\frac{1}{x^3}\right]\,;\;\;\; x,z>\sigma\label{pot_wedge}
 \ee
where the subscript refers to the opening angle $\psi$.  Notice that far from the apex, $x\to\infty$ or $z\to\infty$, this potential reduces to a pure power-law
characteristic of the planar wall (\ref{pot_pla}). The potential (\ref{pot_wedge}) was used in our previous studies of wedge filling \cite{our_prl, our_wedge}. Here,
we extend this analysis to more general wedges whose potential $V_\psi(x,z)$ can be readily obtained by integrating the pair potential $\phi_w(r)$ over a volume of
triangular cross-section which is added either to  $V_\pi(x,z)$ or to $V_{\pi/2}(x,z)$. For \emph{acute} wedges, with $\psi<\pi/2$ (see Fig.~1c), the attractive part
of the potential is
 \bb
V_\psi(x,z)=\alpha_w\left[\frac{1}{z^3}
+\frac{\mathrm{cosec}^3\psi}{(x-z\cot\psi)^3}+\frac{6x^2z^2\cot^2\psi+3z^4\cot^2\psi-6x^3z\cot\psi+2z^4+x^2z^2+2x^4}{2(x-z\cot\psi)^3 z^3\sqrt{x^2+z^2}}\right]\,,
\label{pot_acute}
 \ee
%where the hard wall repulsion now means that $V_\psi(x,z)=\infty$ for $x<\sigma\;{\rm or}\;z>x\tan\psi-\sigma\sec\psi$
with the hard wall repulsion applying within the wall and a distance $\sigma$ from it.  It is clear that $V_\psi(x,z)$ reduces immediately to the expression of
$V_{\pi/2}(x,z)$ for $\psi=\pi/2$.

%\rho_w\int_0^\infty\dd z'\int_0^{z'\cot\psi}\dd x'\int_{-\infty}^\infty\dd y'\phi_w\left(\sqrt{(x-x')^2+y'^2+(z-z')^2}\right)\\

For \emph{open} wedges (see Fig.~1d), corresponding to $\psi>\pi/2$, on the other hand, the potential is more conveniently written as
 \bb
V_{\psi}(x,z)
 =\alpha_w\left[-\frac{6x^2z^2\tan^2\psi+3x^4\tan^2\psi-6z^3x\tan\psi+2x^4+x^2z^2+2z^4}{2(z-x\tan\psi)^3x^3\sqrt{x^2+z^2}} -\frac{\sec^3\psi}{(z-x\tan\psi)^3}
+\frac{2x^4+x^2z^2+2z^4}{2x^3z^3\sqrt{x^2+z^2}}+\frac{1}{z^3}\right]\,,
 \ee
together with the appropriate hard wall restriction. It is easy to verify that this recovers the potential for the right angle corner when  $\psi=\pi/2$ and also the
planar wall when $\psi=\pi$.

%&=&\rho_w\int_{-\infty}^{0}\dd x'\int_{-\infty}^\infty\dd y'\int_0^{-x'\tan\phi'}\dd z'\phi'_w\left(\sqrt{(x-x')^2+y'^2+(z-z')^2}\right)\\

The grand potential functional $\Omega[\rho]$ is minimized numerically on a two dimensional Cartesian square mesh of grid size $0.1\,\sigma$ with appropriate boundary
conditions. We first determine the equilibrium profile for a planar wall $\rho_\pi(z)$  at temperature $T$ and chemical potential $\mu$. This one dimensional density
profile is than imposed as a boundary condition on the two dimensional density along the normals ${\bf n}_1$ and ${\bf n}_2$ at a distance $L=40\,\sigma$ from the
apex along each wall. Previous studies of filling at a right angle corner have shown that distance $L$ is large enough to avoid significant finite size effects and
mimic the interface between the wall and the reservoir fluid.

\begin{figure}[h]
\includegraphics[width=8cm]{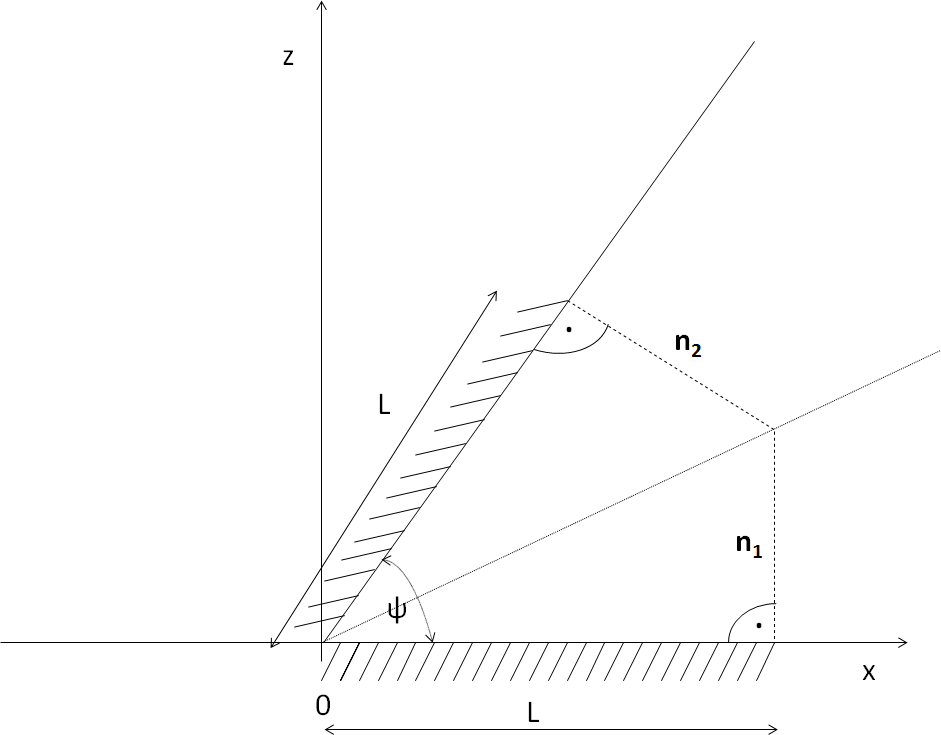}  %\includegraphics[width=8cm]{shallow_wedge}
\caption{Illustration of the finite-size domain used in the numerical minimization of the grand potential functional in acute and open wedges. Along the normals ${\bf
n}_1$ and ${\bf n}_2$ the density is fixed to that of the planar wall-fluid interface, $\rho_\pi(z)$, to mimic the interface with a bulk vapour. The distance $L$ is
set at $L=40\sigma$ which is much larger than the wetting film thickness at a planar wall.} \label{wedges_scheme}
\end{figure}

\section{Results}

We work at two phase bulk coexistence and at subcritical temperatures $T<T_c$ which, for our truncated Lennard-Jones-like potential, occurs at
$k_BT_c/\varepsilon=1.41$. Throughout this work we consider two wall-fluid interaction strengths corresponding to i) $\varepsilon_w=0.8\varepsilon$ and ii)
$\varepsilon_w=0.9\varepsilon$.  We begin by considering the wetting properties of each planar wall-fluid interface determining the temperature dependence of the
contact angle $\theta(T)$, the wetting temperature $T_w$ and, from the numerically determined binding potential, the order of the wetting transition. For both values
of $\varepsilon_w$ the transition is unequivocally first-order. This is to be expected since the wall-fluid interaction is long-ranged while the fluid-fluid potential
is effectively short-ranged. We then consider a right-angle wedge and, from determination of the free-energy and adsorption, locate the filling transition
temperatures $T_f$ which are shown to be completely consistent with the thermodynamic result (\ref{fill}). However the filling transitions are now of different order;
the transition for the weaker potential, for which $T_w$ and $T_f$ are closer to $T_c$ is continuous, in contrast to the stronger potential for which the transition
is first-order. We then investigate what happens to the location and order of these transitions when the opening angle $\psi$ is varied. In particular for each value
of $\varepsilon_w$ we determine the value of the opening angle at which the order of the filling transition changes. Finally, from a numerically constructed wedge
binding potential we are able to determine whether the change in order occurs via a tricritical or critical-end point.

\subsection{The planar wall-fluid interface}

\begin{figure}[h]
\includegraphics[width=10cm]{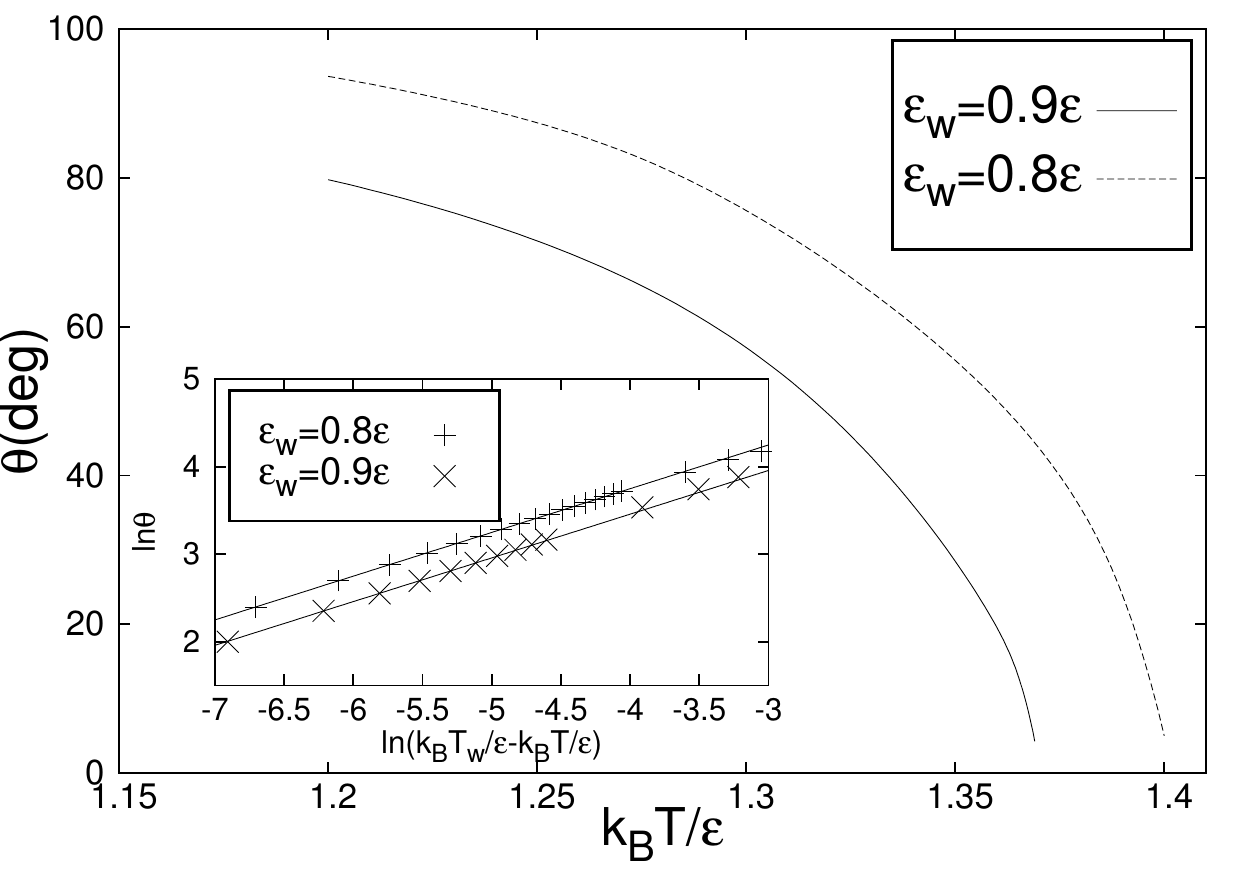}
\caption{Temperature dependence of the contact angle for the wall strengths $\varepsilon_w=0.9\varepsilon$ and $\varepsilon_w=0.8\varepsilon$ showing wetting
transitions at  $k_bT_w=1.37\varepsilon$and $k_bT_w=1.4\varepsilon$ for respectively. In the inset is shown a log-log plot illustrating the vanishing of $\theta$ in
the vicinity of each $T_w$. This is consistent with the expected first-order singularity $\theta(T)\sim (T_w-T)^{\frac{1}{2}}$ -- the straight lines have slope equal
to $1/2$. From the temperature dependence of the contact angle one may also read off the macroscopic prediction for the location of the filling temperature in wedges
of opening angle $\psi$ according to $\theta(T_f)=(\pi-\psi)/2$.
 }\label{cont_ang}
\end{figure}

\begin{figure}[h]
\includegraphics[width=10cm]{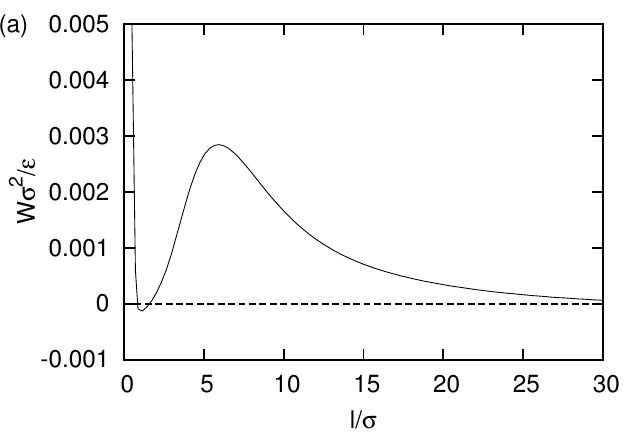}
\includegraphics[width=10cm]{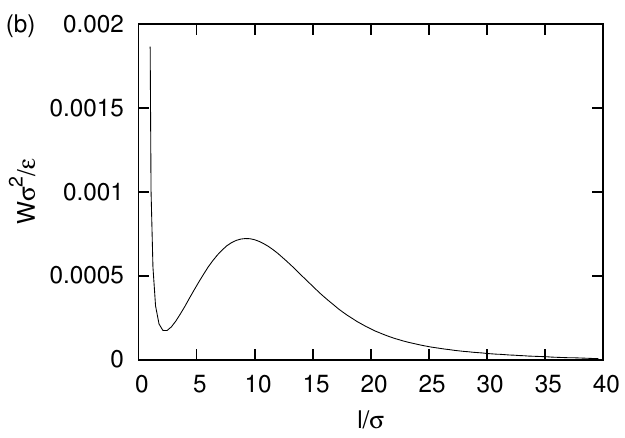}
\caption{Numerically determined binding potential functions for the planar wall-fluid interface. The upper panel (a) refers to wall strength
$\varepsilon_w=0.9\varepsilon$ and $T<T_w$, while the bottom panel (b) refers to $\varepsilon_w=0.8\varepsilon$ and $T>T_w$. The presence of the potential barrier
indicates that the wetting transition is of first-order in both cases but note that the barrier is an order of magnitude smaller for $\varepsilon_w=0.8\varepsilon$
consistent with $T_w$ being much closer to $T_c$.}\label{bind_plane}
\end{figure}

We consider first a planar wall of infinite area $A$, with the power-law potential (\ref{pot_pla}), in contact with a bulk fluid of volume $V$ at pressure $p$.
Imposing that $\rho(\infty)= \rho_l$ or $\rho(\infty)= \rho_g$ (corresponding to the bulk liquid and gas densities respectively) determines the equilibrium profiles
$\rho(z)$ and surface tensions $\gamma=(\Omega+pV)/A$ for the wall-liquid (wg) and wall-gas (wg) interfaces respectively. Then from Young's equation
$\gamma_{wg}=\gamma_{wl}+\cos\theta\gamma_{lg}$, where $\gamma_{lg}$ is the liquid-gas interfacial tension, we determine the temperature dependence of the macroscopic
contact angle $\theta(T)$. This is shown in Fig.~\ref{cont_ang}, and indicates the presence of a wetting transition occurring at $k_bT_w=1.37\varepsilon$ for
$\varepsilon_w=0.9\varepsilon$ and $k_bT_w=1.4\varepsilon$ for $\varepsilon_w=0.8\varepsilon$. The asymptotic behaviour as $T\to T_w$ is consistent with
$\theta(T)\sim (T_w-T)^{\frac{1}{2}}$, indicating that the wetting transitions are both first-order, as expected (see inset). As a further check on this we have
numerically determined the binding potential $W(\ell)=\Omega(\ell)/A-\gamma_{wl}-\gamma_{lg}$, where $\Omega(\ell)$ is the grand potential of a wetting layer
constrained to be of thickness $\ell$ (see Fig.~\ref{bind_plane}). These show an activation barrier between the partially wet and completely wet states close to $T_w$
confirming the first-order nature of the transitions. More generally, from the plots for $\theta(T)$ we can now test the thermodynamic prediction (\ref{fill}) for the
location of the filling transition in wedges with different opening angles, beginning with the right angle corner.

\subsection{The right angle wedge}

\begin{figure}
\includegraphics[width=10cm]{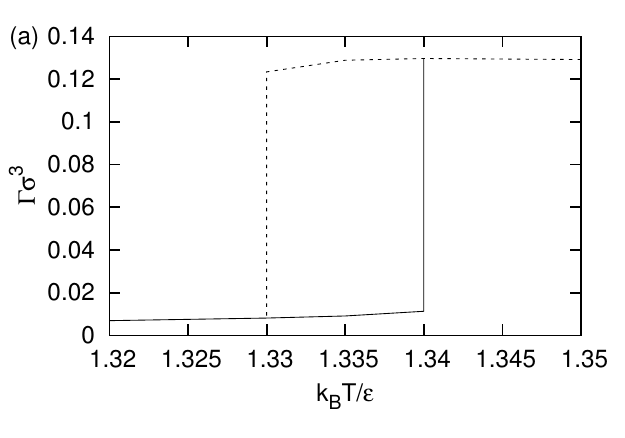}
\includegraphics[width=10cm]{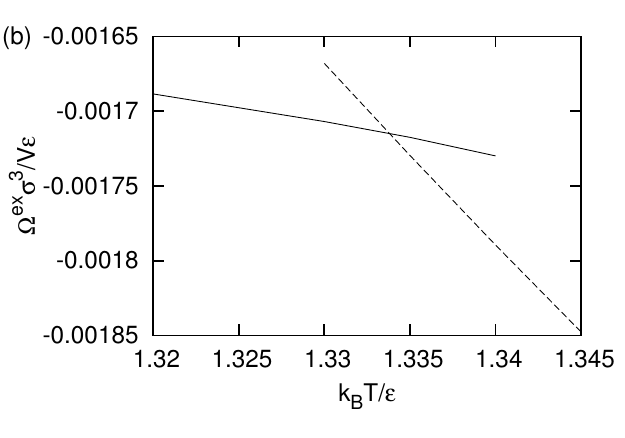}
\caption{Adsorption (a) and excess grand potential density (b) as a function of temperature for a rectangular wedge with wall strength $\varepsilon_w=0.9\varepsilon$.
Two distinct branches are obtained by minimization of $\Omega[\rho]$ from initial low-density high-density configurations respectively. The branches in the grand
potential cross at $k_BT=1.335\,\varepsilon$ consistent with the thermodynamic prediction obtained from Eqn.~(1). The hysteresis is indicative of a first-order
filling transition. } \label{rec_ew09}
\end{figure}

\begin{figure}[h]
\includegraphics[width=7cm]{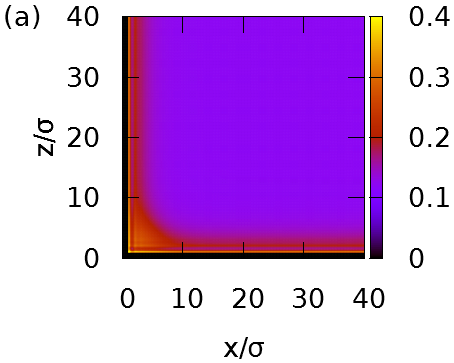} \hspace*{0.5cm} \includegraphics[width=7cm]{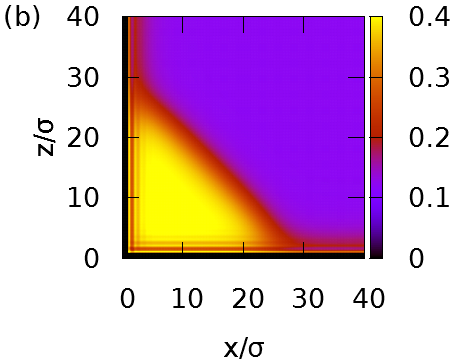}
\caption{Coexisting equilibrium density profiles for a rectangular wedge with the wall strength $\varepsilon_w=0.9\varepsilon$ at the filling temperature
$k_BT_f=1.335\varepsilon$. The low-density state (a) corresponds to the adsorption (solid) lines in Fig.~5 and the high density state (b) corresponds to the
desorption (dashed) lines in Fig.~5. In the high-density state, the liquid vapour interface meets the wall at the contact angle $\theta=\pi/4$ in agreement with the
macroscopic prediction of Eq.~(\ref{fill}).
 }\label{profs_90}
\end{figure}

\begin{figure}
\includegraphics[width=10cm]{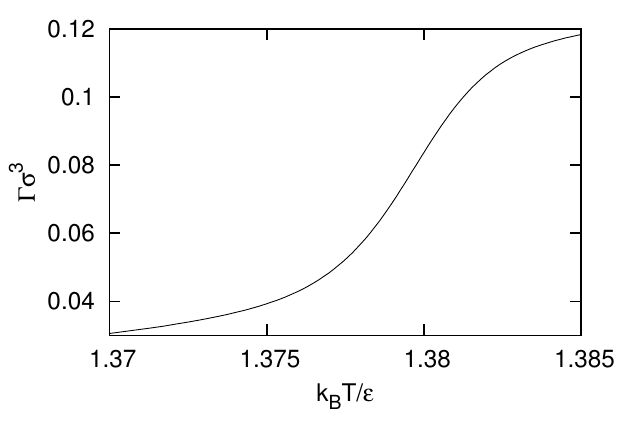}
\caption{Temperature dependence of the adsorption in a rectangular wedge with wall strength $\varepsilon_w=0.8\varepsilon$. The curve has its maximum gradient at $k_B
T/\varepsilon\approx 1.38$ very close to the location of the filling transition $k_BT_f\approx1.378\varepsilon$ predicted by Eqn.~(\ref{fill}). There is no hysteresis
indicating the filling transition is continuous.} \label{rec_ew08}
\end{figure}

%\begin{figure}[ht]
%\includegraphics[width=5cm]{rec_ew08.eps}
%\caption{ }\label{rec_ew08}
%\end{figure}

We now consider the filling transitions in a right-angle wedge, corresponding to the potential $V_{\pi/2}(x,z)$, for the two different values of $\varepsilon_w$.
Using the plots of $\theta(T)$, the macroscopic result (\ref{fill}), predicts that $k_BT_f/\epsilon \approx 1.33 $ for $\varepsilon_w=0.9\varepsilon$ and
$k_BT_f/\epsilon \approx1.375$ for $\varepsilon_w=0.8\varepsilon$. First consider the excess adsorption, defined for the wedge-gas interface, by
 \bb
 \Gamma=\frac{1}{L^2}\int\dd x\int\dd z\left[\rho(x,z)-\rho_g\right]\,.\label{ads}
 \ee
For large enough coverage, i.e., for mesoscopicaly large values of the excess adsorption, $\Gamma\propto \ell_w^2(\rho_l-\rho_g)/L^2$ where $\ell_w$ is the
perpendicular distance of the liquid-vapour interface from the wedge apex. the stronger wall-fluid potential, $\varepsilon_w=0.9$ the adsorption shows a jump, close
to the predicted value of $T_f$, between two states corresponding to small and large coverage of liquid near the wedge apex. This is reflected in the temperature
dependence of the grand-potential which shows that two branches cross at $T_f=1.335$ which is in excellent agreement with (\ref{fill}) (see Fig.~\ref{rec_ew09}). The
coexisting profiles are shown in Fig.~\ref{profs_90}. Note that in the higher coverage state the interface is essentially flat and meets the walls at the correct
contact angle $\theta(T_f)\approx \pi/4$. The coverage of this state scales with $L^2$ corresponding, in the thermodynamic limit, to a completely filled wedge. For
the weaker wall potential however there is no coexistence and the adsorption (and grand potential) has a single branch which rapidly, but smoothly, increases as the
temperature is raised towards $T_f$ (see Fig.~\ref{rec_ew08}). This indicates that the filling transition is continuous despite the fact that the wetting transition
is first-order.

\subsection{Non-rectangular wedges}

\begin{figure}[h]
\includegraphics[width=10cm]{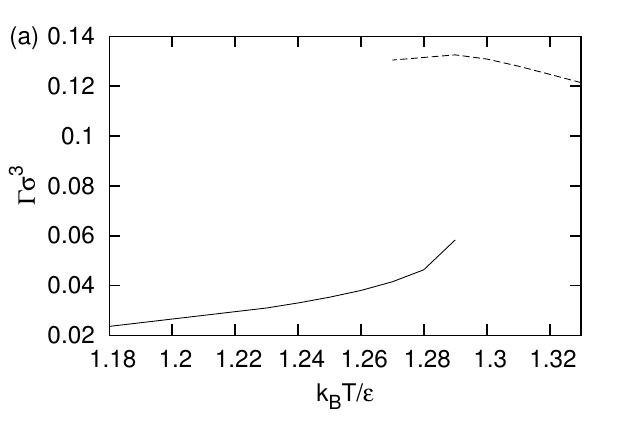}
\includegraphics[width=10cm]{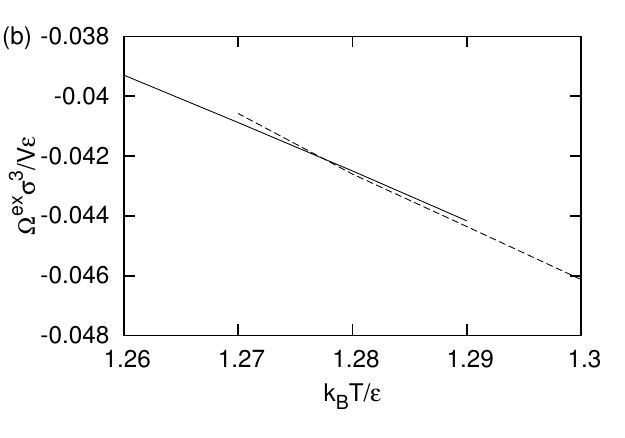}
\caption{Adsorption (a) and excess grand potential density (b) as a function of temperature for a wedge with opening angle $\psi=45\degree$ and wall strength
$\varepsilon_w=0.9\varepsilon$. The filling transition is located at $k_BT_f\approx1.28\varepsilon$ and is weakly first-order.
 }\label{ads_45}
\end{figure}

\begin{figure}[h]
\includegraphics[width=7cm]{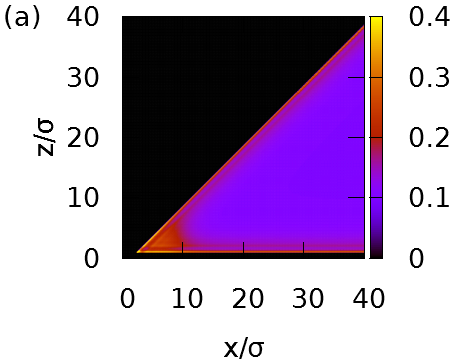}\hspace*{0.5cm} \includegraphics[width=7cm]{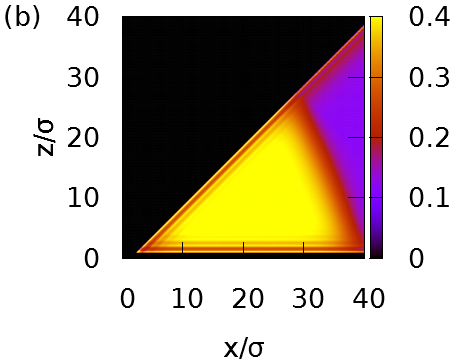}
\caption{(Color online) Coexisting equilibrium density profiles for a wedge with opening angle $\psi=45\degree$ and wall strength $\varepsilon_w=0.9\varepsilon$ at
the filling temperature $k_BT_f=1.28\varepsilon$.
 }\label{profs_45}
\end{figure}

\begin{figure}[h]
\includegraphics[width=10cm]{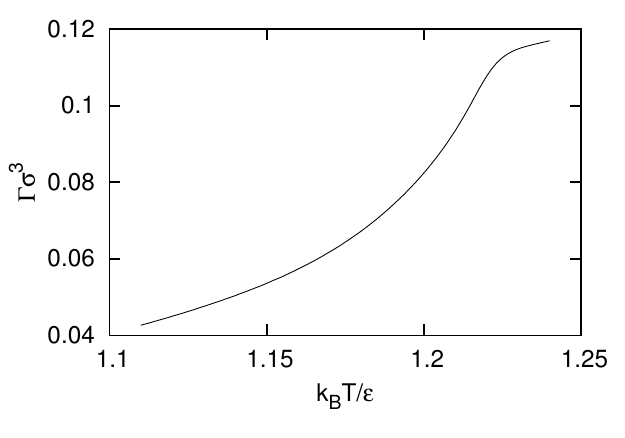}
\caption{Temperature dependence of the adsorption for a wedge with opening angle $\psi=20\degree$ for wall strength $\varepsilon_w=0.9\varepsilon$. The filling
transition at $k_BT_f\approx1.22\varepsilon$ is continuous.
 }\label{ads_20}
\end{figure}

\begin{figure}[h]
\includegraphics[width=10cm]{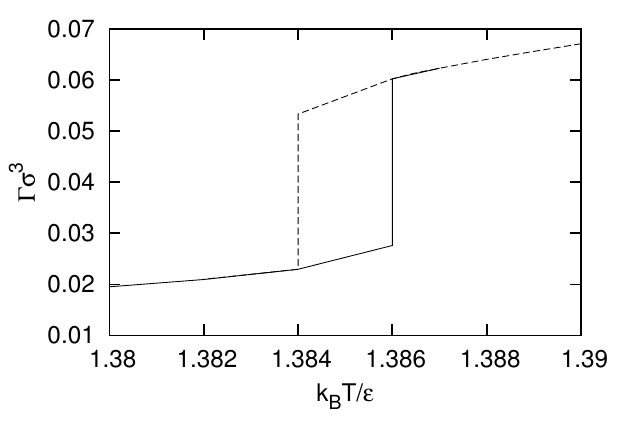}
\caption{Hysteresis in the adsorption as a function of temperature for a wedge, with opening angle $\psi=120\degree$, and wall strength
$\varepsilon_w=0.8\varepsilon$.
 }\label{ads_120}
\end{figure}

\begin{figure}[h]
\includegraphics[width=7cm]{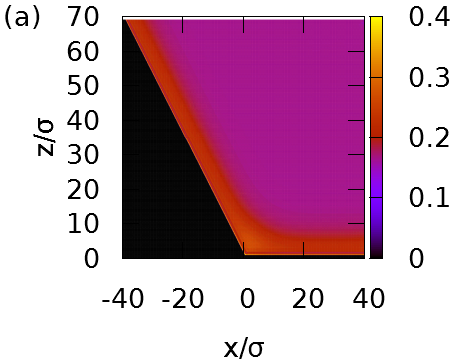}\hspace*{0.5cm}\hspace*{1cm}\includegraphics[width=7cm]{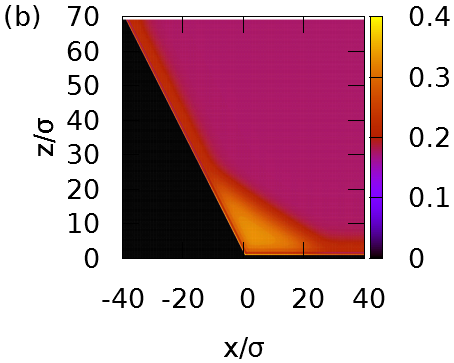}
\caption{(Color online) Coexisting equilibrium density profiles for a wedge with opening angle $\psi=120\degree$ and wall strength $\varepsilon_w=0.8\varepsilon$ at
the filling temperature $k_BT_f=1.385\varepsilon$.
 }\label{profs_120}
\end{figure}

Consider first the stronger wall, $\varepsilon_w=0.9\varepsilon$, for which the filling transition is first-order for the right-angle wedge, $\psi=\pi/2$. We now
close the wedge so that the opening angle $\psi=45$ and repeat our analysis, determining the temperature dependence of the adsorption and grand potential after
minimizing $\Omega[\rho]$ from low-density and high-density initial configurations. Once again we find hysteresis in the adsorption but this is now much diminished
indicating the transition is only weakly first-order. Correspondingly there are two branches to the grand potential but these now meet almost tangentially at $k_B
T_f/\varepsilon=1.28$. Again this value of $T_f$ is in full agreement with the macroscopic prediction (\ref{fill}). The coexisting states at the filling temperature
are displayed in Fig.~\ref{profs_45}. Note that the small decrease in the adsorption with increasing $T$ in the high coverage phase is simply due to decrease in the
$\rho_l-\rho_g$ and does not reflect the position $\ell_w$ of the interface which is saturated. Decreasing the opening angle further reduces the hysteresis which
eventually vanishes when the wedge is very acute, when $\psi\approx 20$. This is illustrated in Fig.~\ref{ads_20} which shows only a smooth but rapid increase of the
adsorption as $T\to T_f$ where $k_B T_f/\varepsilon\approx 1.2$. In this highly acute wedge the filling transition has therefore become continuous.

The opposite happens for the weaker potential $\varepsilon_w=0.8$ for which the filling transition was already continuous for the right angle wedge. In this case,
opening the wedge eventually induces hysteresis when $\psi\approx 120\degree$ (see Fig.~\ref{ads_120}) indicating that the filling transition, like the underlying
wetting transition, is first-order. For $\psi=120\degree$ the crossing of the branches of the grand potential determines $k_BT_f=1.385\varepsilon$ which is in perfect
agreement with (\ref{fill}). The coexisting states at $T_f$ are shown in Fig.~\ref{profs_120}.

\section{Discussion and concluding remarks}

In this work, we have presented a non-local density functional study of filling transitions in open and acute wedges, extending previous studies which were restricted
to right-angle corners. In our model the walls of the wedge themselves exhibit a wetting transition (at temperature $T_w$) which is always first-order in nature
regardless of the strength of the wall-fluid interaction $\varepsilon_w$. We have found that in the wedge geometry, the location of the filling transition temperature
$T_f$ is always in agreement with the thermodynamic prediction, $\theta(T_f)=(\pi-\psi)/2$, indicating that $T_f$ can be arbitrarily lowered below $T_w$ by decreasing
the opening angle $\psi$. In addition we found that by reducing the opening angle one can always drive the filling transition second-order implying that the
adsorption continuously changes from micro- to macroscopic at $T_f$. This generalizes our earlier studies of filling at right angle corners and shows that the change
in order is not restricted to transitions in the proximity of $T_c$.

\begin{figure}[h]
\includegraphics[width=10cm]{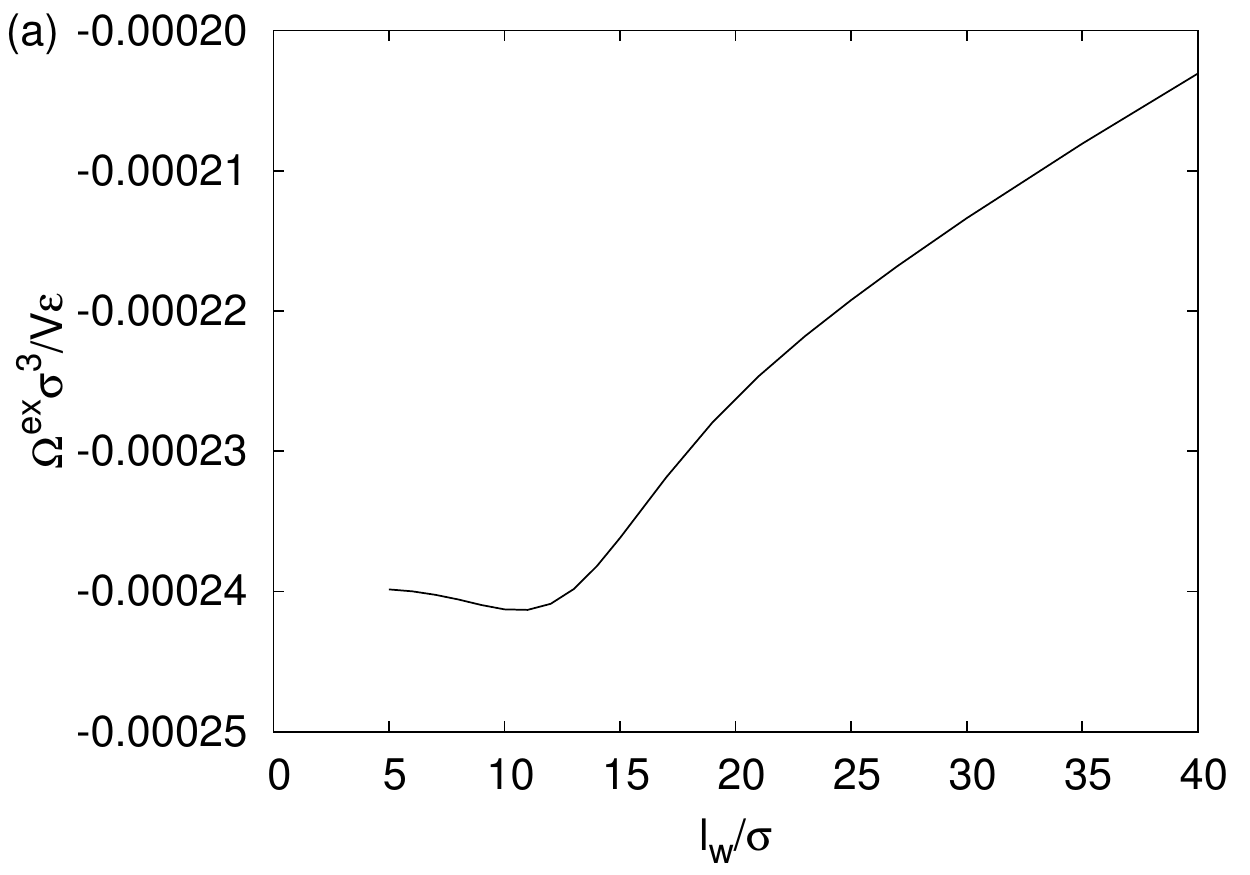} \hspace*{0.2cm}\includegraphics[width=10cm]{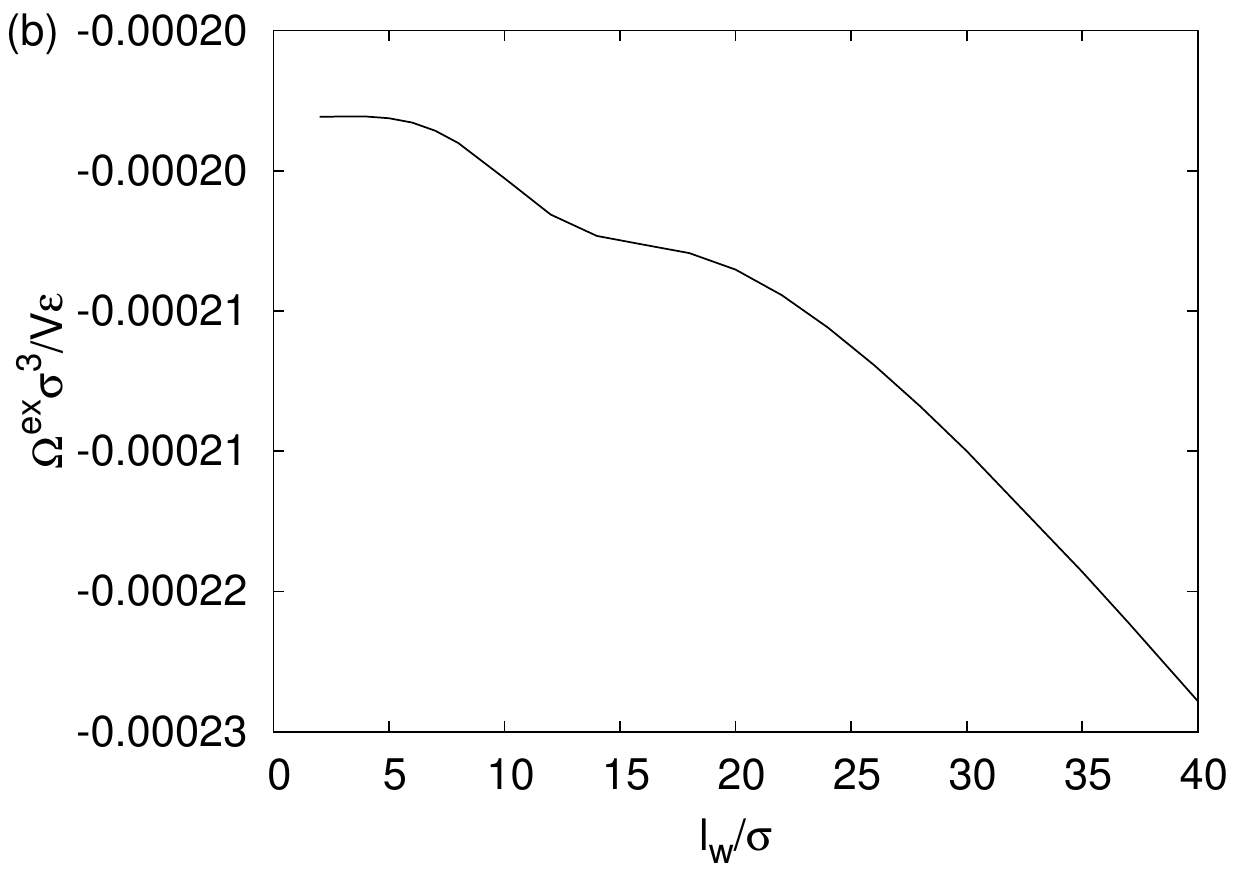}
\caption{Constrained grand potential as a function of the film thickness $\ell_w$ for a right-angle wedge with the wall strength $\varepsilon_w=0.8\varepsilon$. For
temperatures below the filling temperature $k_BT_f=1.378\varepsilon$ (upper panel) the potential exhibits a single minimum which continuously shifts away from the
wedge apex as the temperature increases. In the bottom panel the tail of the binding potential has a negative slope as required since $T>T_f$. There is no potential
barrier between a local minimum and a global extremum, indicating the change in order is via tricritical point rather than a critical end-point.
 }\label{bp08}
\end{figure}

Our central finding, that it is possible to induce continuous filling for sufficiently acute wedges, is in {\it{partial}} agreement with longstanding predictions of
simple effective Hamiltonian theory for filling in open wedges (with $\psi\approx\pi$) \cite{wood2}. Such interfacial models also predict a change in order from first
to continuous filling,  when the filling temperature $T_f$ is sufficiently below $T_w$ that there is no activation barrier in the binding potential $W(\ell)$, defined
for wetting at the planar wall. Within such models this means that for wedges made from walls that only exhibit first-order wetting, the filling transition is
first-order when the wedge is open, but continuous when it is sufficiently acute. This is in qualitative agreement with our findings. However the mechanism behind the
change of order cannot be exactly the same as that within the interfacial Hamiltonian description. This is because within our present study only the wall-fluid
potential is long-ranged which means that binding potential for wetting at a planar wall always has an activation barrier. The activation barrier would not be present
if one could induce a  change in the sign of the leading order term in the binding potential (Hamaker constant) which in turn needs a balance between the strengths of
long-ranged wall-fluid and long-ranged fluid-fluid potentials. Therefore, if one strictly applied the predictions of the interfacial Hamiltonian model to the present
system then the filling transition would always be first-order albeit very weak since the strength of the activation barrier is rather small at  $T_f$.

So what is the reason for this discrepancy? One option is that the finite-size restrictions in the present numerical study have rounded the filling phase transition,
which for larger domain sizes $L$ would be (weakly) first-order. While we cannot completely rule out this possibility, it is notable that the numerically determined
location of the filling transition is always in excellent agreement with the thermodynamic result $\theta(T_f)=\alpha$. The location of the transition therefore is
certainly not strongly influenced by the finite-size. As a check on this we have repeated our analysis of filling a rectangular corner for
$\varepsilon_w=0.8\,\varepsilon$ for the much large domain size up to $L=100\sigma$ which again shows only a smooth increase in the adsorption consistent with
continuous filling. The second option, which appears more likely to us, is that the original effective Hamiltonian description does not capture all the details of the
filling transition. There are indeed plausible reasons for believing this since the original interfacial Hamiltonian model is only applicable to shallow wedges
{\it{and}} to filling temperatures $T_f$ far from $T_c$ where a simple sharp-kink description of the interface structure is reliable. If the wedge is very acute or if
$T_f\approx T_c$ then a sharp-kink approximation ceases to be valid. However these are precisely the conditions where we find a change in the order of the filling
transition. Extending the original effective Hamiltonian model of filling to these regimes requires, at least, both a soft-kink treatment of the non-planar interface
and a fully non-local description of the interface-wall potential. In addition in acute wedges, packing and volume exclusion effects are almost certain to play an
important role and are clearly visible in the density profile. For continuous and weakly first-order filling transitions, the free-energy landscape, determining for
example the energy cost of maintaining a coverage of order $\Gamma \propto \ell_w^2$ is so shallow, that any extra stress on the liquid-vapour interface may strongly
effect the phase transition. Given that packing effects are completely neglected in the original effective Hamiltonian theory it appears to us highly likely that this
is the source of any new physics within the microscopic density functional description of filling. Incorporating all these features into the interfacial Hamiltonian
theory is extremely challenging, indeed so much so, that a microscopic density functional treatment is a much more tractable way of studying the problem. Our results
suggest that further work is required to understand how packing effects can lead to extra terms in the binding potential for wedge filling which may compete with
those arising directly from the intermolecular forces.

A very subtle question which we have not yet addressed concerns the precise nature of the change in order of the transition. In principle this may happen via one of
two mechanisms: a tricritical point or a critical-end point. If there was a critical end-point then in the range of $\psi$ values where the filling transition is
continuous there would still be a meta-stable low coverage state even at the filling temperature $T_f$. In Fig.~\ref{bp08} we show plots of the numerically determined
grand potential, obtained via partial minimization, as a function of a constrained value $\ell_w$ of the thickness of liquid from the wedge apex, in the second-order
filling regime. The left panel corresponds to $T$ slightly below $T_f$ while the right hand panel is slightly above $T_f$. Both graphs have a linear contribution
$\propto (T_f-T)l_w$ proportional to the film thickness which changes sign at the filling temperature. This has a purely thermodynamic origin arising from the surface
tensions and is responsible for the macroscopic prediction (\ref{fill}). As the temperature is increased the location of the minimum smoothly increases and eventually
disappears close to $T_f$ when the linear term changes sign. It is clear there is no local minimum when $T>T_f$ indicating that the change in order of the filling
transition is via tricriticality. This means if we were to sit along the line of first-order filling transition temperature $T_f$ and decrease the opening angle
$\psi$, the adsorption of the low coverage phase would diverge continuously as we approach the tricritical value of $\psi$. However studying the nature of this
divergence in more detail would be extremely difficult do to finite-size constraints.

Finally, we mention that our density functional study is mean-field in nature and neglects long wavelength fluctuation effects associated with thermal wandering of
the interface along the wedge. These certainly do not alter the location the filling boundary, $\theta(T_f)=(\pi-\psi)/2$, which is determined  by surface
thermodynamics, nor the underlying mechanism for the change in the order of the phase transition, which depends on the competition between geometry and long-ranged
intermolecular forces. The only influence of thermal fluctuations of any import concerns the roughness $\xi_\perp$ of the liquid-gas interface, which in the regime
where the filling transition is second-order, is expected to diverge according to a universal power-law $\xi_\perp\approx (T_f-T)^{-1/4}$. This is not allowed for in
mean-field density functional studies, which, as is well known, always yields an interfacial width of order the bulk correlation length. In the presence of
long-ranged intermolecular forces however, the roughness $\xi_\perp$, even allowing for interfacial wandering is always much less than the equilibrium film thickness
of liquid adsorbed near the wedge apex, and mean-field predictions for all other quantities of interest should be reliable.

\begin{acknowledgments}
 \noindent A.M. acknowledges the support from the Czech Science Foundation, project 13-09914S. A.O.P. wishes to thank the support of the EPSRC UK
 for Grant No. EP/J009636/1 and EPSRC Mathematics Platform grant EP/I019111/1.
\end{acknowledgments}

\end{document}